\documentclass[letterpaper, 10 pt, conference]{ieeeconf}  

\IEEEoverridecommandlockouts                              
\usepackage{textcomp, gensymb}
\usepackage{booktabs,threeparttable}
\usepackage{lipsum}
\usepackage{hyperref}

\usepackage{graphicx} 
\usepackage{epsfig} 
\usepackage{mathptmx} 
\usepackage{times} 
\usepackage{amsmath} 
\usepackage{amssymb} 
\usepackage[caption=false,font=normalsize,labelfont=sf,textfont=sf]{subfig}
\usepackage{tabularx}

\title{\LARGE \bf
Relocating thermal stimuli to the proximal phalanx may not affect vibrotactile sensitivity on the fingertip
}

\author{Huibert~A.~J.~van~Riessen$^{1}$ and Yasemin~Vardar$^{1}$% <-this % stops a space
\thanks{$^{1}$Authors are with the Department of Cognitive Robotics,  Faculty of  3ME,  Delft  University of  Technology,  2628  CD  Delft, The Netherlands.
        {\tt\small Y.Vardar@tudelft.nl}}%
}

\begin{document}

\maketitle
\thispagestyle{empty}
\pagestyle{empty}

%%%%%%%%%%%%%%%%%%%%%%%%%%%%%%%%%%%%%%%%%%%%%%%%%%%%%%%%%%%%%%%%%%%%%%%%%%%%%%%%
\begin{abstract}
Wearable devices that relocate tactile feedback from fingertips can enable users to interact with their physical world augmented by virtual effects. While studies have shown that relocating same-modality tactile stimuli can influence the one perceived at the fingertip, the interaction of cross-modal tactile stimuli remains unclear. Here, we investigate how thermal cues applied on the index finger's proximal phalanx affect vibrotactile sensitivity at the fingertip of the same finger when employed at varying contact pressures. We designed a novel wearable device that can deliver thermal stimuli at adjustable contact pressures on the proximal phalanx. Utilizing this device, we measured the detection thresholds of fifteen participants for 250~Hz sinusoidal vibration applied on the fingertip while concurrently applying constant cold and warm stimuli at high and low contact pressures to the proximal phalanx. Our results revealed no significant differences in detection thresholds across conditions. These preliminary findings suggest that applying constant thermal stimuli to other skin locations does not affect fingertip vibrotactile sensitivity, possibly due to perceptual adaptation. However, the influence of dynamic multisensory tactile stimuli remains an open question for future research.

\end{abstract}

%%%%%%%%%%%%%%%%%%%%%%%%%%%%%%%%%%%%%%%%%%%%%%%%%%%%%%%%%%%%%%%%%%%%%%%%%%%%%%%%
\section{Introduction}
Wearable haptic interfaces that can display rich arrays of tactile cues play an essential role in bridging the gap between the physical world interactions and those in virtual environments. Most of these devices~\cite{gabardi2016new}~\cite{murakami2017altered}~\cite{spagnoletti2018rendering} embed a variety of sensors and actuators and are assembled in the form of a glove or ring to directly actuate fingertips, as these are the most sensitive skin regions and humans mostly use their fingertips to interact with physical objects. These devices can be programmed to display material sensations in virtual environments through controlled vibration, friction, pressure, or thermal stimuli on the fingertips. 

Despite the advantages of generating naturalistic touch in virtual environments, mounting actuators directly on fingertips also presents disadvantages. For example, they limit the range of motion by introducing additional thickness to the fingertip. This situation increases the likelihood of collisions and could result in the inability to perform tasks featuring small objects or confined spaces. Moreover, occlusion of fingertips prevents direct interaction with the physical environment, limiting the usage of devices for mixed reality applications~\cite{teng2022xr}. Nonetheless, placing actuators directly on the fingertip also causes occlusions for hand tracking performed by computer vision algorithms~\cite{pacchierotti2016hring}. 

\begin{figure} [t]
    \centering
  \subfloat[ ]{%
       \includegraphics[width=0.4\linewidth]{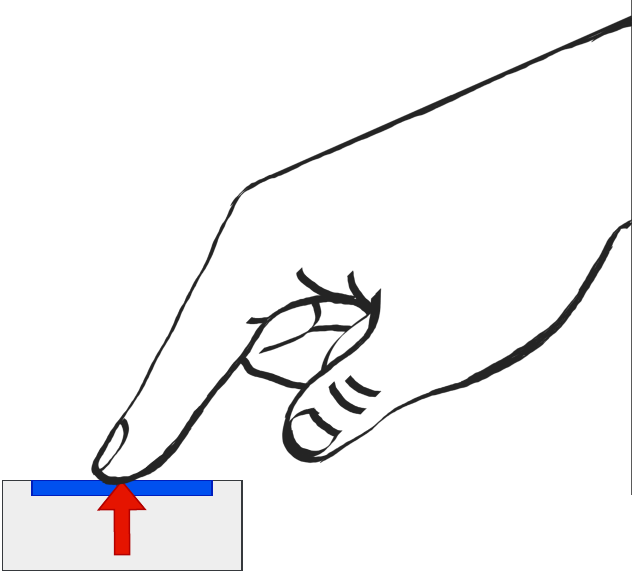}}
    \hfill
  \subfloat[ ]{%
        \includegraphics[width=0.4\linewidth]{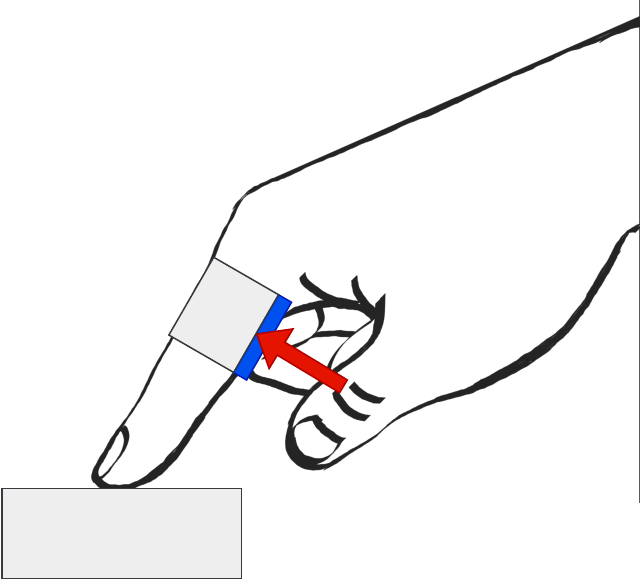}}
  \caption{(a) Thermal (blue) stimulus perceived simultaneously by interacting with a material surface at applied contact pressure (red). (b) Relocated thermal (blue) stimulus is applied to the proximal phalanx with a pressure (red) and simultaneously perceived during interaction with the material surface using the fingertip.}
  \label{fig_relocation} 
\end{figure}

Recent works proposed relocating tactile sensations to other body parts to free the fingertip and prevent the above problems. Pacchierotti et al.~\cite{pacchierotti2016hring} introduced the ”hRing” device to reduce occlusions in hand tracking sensors. hRing had two servo motors in combination with a belt to render pressure and skin stretch at the proximal phalanx of the index finger. They demonstrated that displaying cutaneous feedback via this device helped users pick and place virtual objects. Moreover, Gioioso et al.~\cite{gioioso2018using} showed that applying thermal and pressure cues to the proximal phalanx with another ring-type device could enable users to discriminate the temperature of objects in virtual reality when interacting with them with their hands, even though the thermal cues were not directly applied there. Palmer et al.~\cite{palmer2022haptic} proposed mappings for relocating forces from the thumb and index finger to the wrist and demonstrated their benefits during pick-and-place tasks when visual feedback is limited. Pezent et al.~\cite{pezent2022explorations} proposed "Tasbi", a wrist-worn device that combines vibrotactile and squeeze stimuli during augmented and virtual reality interactions, and showed multiple applications where these relocated tactile stimuli are combined with visual feedback to create visual pseudo-haptic illusions of tactile interactions~\cite{pezent2023multisensory}. Tanaka et al.~\cite{tanaka2023full} introduced a system in which they attached electrodes to the wrist and the back of the hand to apply electro-tactile stimuli to the nerves while keeping the palmar side unobstructed. Their experiments showed that participants most dominantly localized the tactile sensations on the palmar side of the hand while they were applied to the dorsal side. Other studies demonstrated that relocating vibration stimuli on the proximal phalanx of the index finger~\cite{friesen2023perceived} or wrist~\cite{gaudeniRelocate} can also be utilized for displaying information, such as the texture of materials. 

Apart from generating tactile cues to manipulate or feel virtual objects, placing actuators away from fingertips also allows altering sensations of physical surroundings perceived with fingertips, opening new opportunities for mixed reality applications when used in a controlled manner. For example, previous research~\cite{de2018enhancing, salazar2020altering} showed that relocating pressure and friction stimuli via ”hRing” can alter the perceived stiffness of tangible objects or give the feeling of a bump or hole on a flat surface. Asano et al.~\cite{asano2014vibrotactile} demonstrated that a vibrating voice-coil actuator worn on the side of the finger can modify the perceived roughness of physical objects. Moreover, Jamalzadeh et al.~\cite{jamalzadeh2019effect} showed that subthreshold vibrotactile stimuli applied on the proximal phalanx of the index finger increased the detection threshold of electrovibration stimuli perceived on the fingertip. 

The mentioned examples prove the feasibility of relocating tactile stimuli with a perceptual modality for altering a \emph{related} perceived sensation of an object (e.g., remote pressure vs. sensed stiffness). However, it is unclear how tactile stimuli perceived on the fingertip would be affected by relocated tactile stimuli with a different modality.

Prior research provided compelling evidence of perceptual interaction between thermal and tactile stimuli when applied at the \emph{same} skin site. For instance, Green~\cite{green1977effect} demonstrated that sensitivity of the thenar eminence to 250 Hz vibration is highest at 34~\degree and decreases substantially by cooling the skin to 20~\degree C. Later, Klinenberg et al.~\cite{klinenberg1996temperature} repeated similar experiments on the fingertip and confirmed the decline in vibrotactile sensitivity at 17~\degree C. Subsequent studies by Hirosawa et al. and Burstrom et al. demonstrated that interaction is bidirectional; they reported a reduction in sensitivity for warm~\cite{hirosawa1992thermal} and cold~\cite{burstrom2009thermal} stimuli due to exposure to vibration on the hand. Singhal and Jones~\cite{singhal2017interaction} also showed that concurrent vibrotactile stimulation can influence the ability to identify thermal patterns. Interestingly, a subsequent study by Park et al.~\cite{park2022preliminary} observed independence between these two modalities during information transfer. Other researchers also reported perceptual interactions between thermal stimuli on other modalities, such as pressure sensation~\cite{weber1996tactile, stevens1977pressure}, tactile acuity~\cite{stevens1982acuity}, and roughness~\cite{green1979roughness}. 

Recent research has also demonstrated perceptual interactions between thermal and tactile stimuli applied to \emph{different} skin sites. Liu et al.~\cite{liu2021thermo} showed that the perceived location of a thermal stimulus can be systematically shifted towards the location of a simultaneous pressure stimulus applied to a nearby position on the arm. The authors attributed this phenomenon to an extension of thermal referral, which refers to the perceived location of thermal stimuli shifted to a nearby location. Subsequently, Son et al.~\cite{son2023thermo} observed a similar behavior with vibration stimuli applied to the back. They demonstrated that high-intensity vibrations can induce localized thermal sensations away from the source. They explained this phenomenon as perceptual masking, occurring due to the high-intensity vibration overriding the perception of the thermal stimulus at the source. 

Inspired by prior research on cross-modal sensory interaction, we investigated the effect of remotely applied thermal stimuli under varying pressures on vibrotactile sensitivity at the fingertip. We hypothesized that simultaneous exposure to strong (non-noxious) warm or cold stimuli to the proximal phalanx would elevate vibrotactile thresholds due to perceptual masking~\cite{son2023thermo}. Considering that applied contact force also affects contact area and skin thermal resistivity, we anticipated a positive correlation between the degree of masking and the contact pressure of the applied thermal stimulus. If remote thermal stimuli would impact vibrotactile sensitivity at the fingertip, careful consideration is warranted when designing tactile stimuli. For instance, reduced vibrotactile sensitivity could limit the range of perceivable roughness in digital textures~\cite{yoshioka2007texture} or hinder performance during manual tasks~\cite{lockhart1975effect}. Conversely, if these multimodal stimuli are processed independently, positioning actuators away from fingertips could benefit designing haptic displays for tasks that demand fingertip dexterity and multisensory feedback (e.g., robotic surgery).

To answer our research question, we first developed a novel wearable haptic device, which could deliver thermal and pressure stimuli to the proximal phalanx of the index finger and vibration stimuli to the fingertip; see Fig.~\ref{fig_relocation} for an illustration. Then, we measured the detection thresholds of fifteen participants for a vibrotactile stimulus applied to their fingertips while perceiving a set of simultaneous thermal and pressure stimuli on the proximal phalanx. 

\section{Methods}
\subsection{Participants}
Twelve males and three females, all right-handed, between the ages of 21 and 32 (mean: 24.5, standard deviation: 2.7), participated in the experiments. None of them had injuries or neurological problems in their right hands. The experimental procedures were conducted following the Declaration of Helsinki and approved by the Human Research Ethics Committee of TU Delft with case number 3279. All participants gave informed consent. 

\subsection{Experimental setup}
\begin{figure} 
    \centering
      \subfloat[ ]{%
        \includegraphics[width=\linewidth]{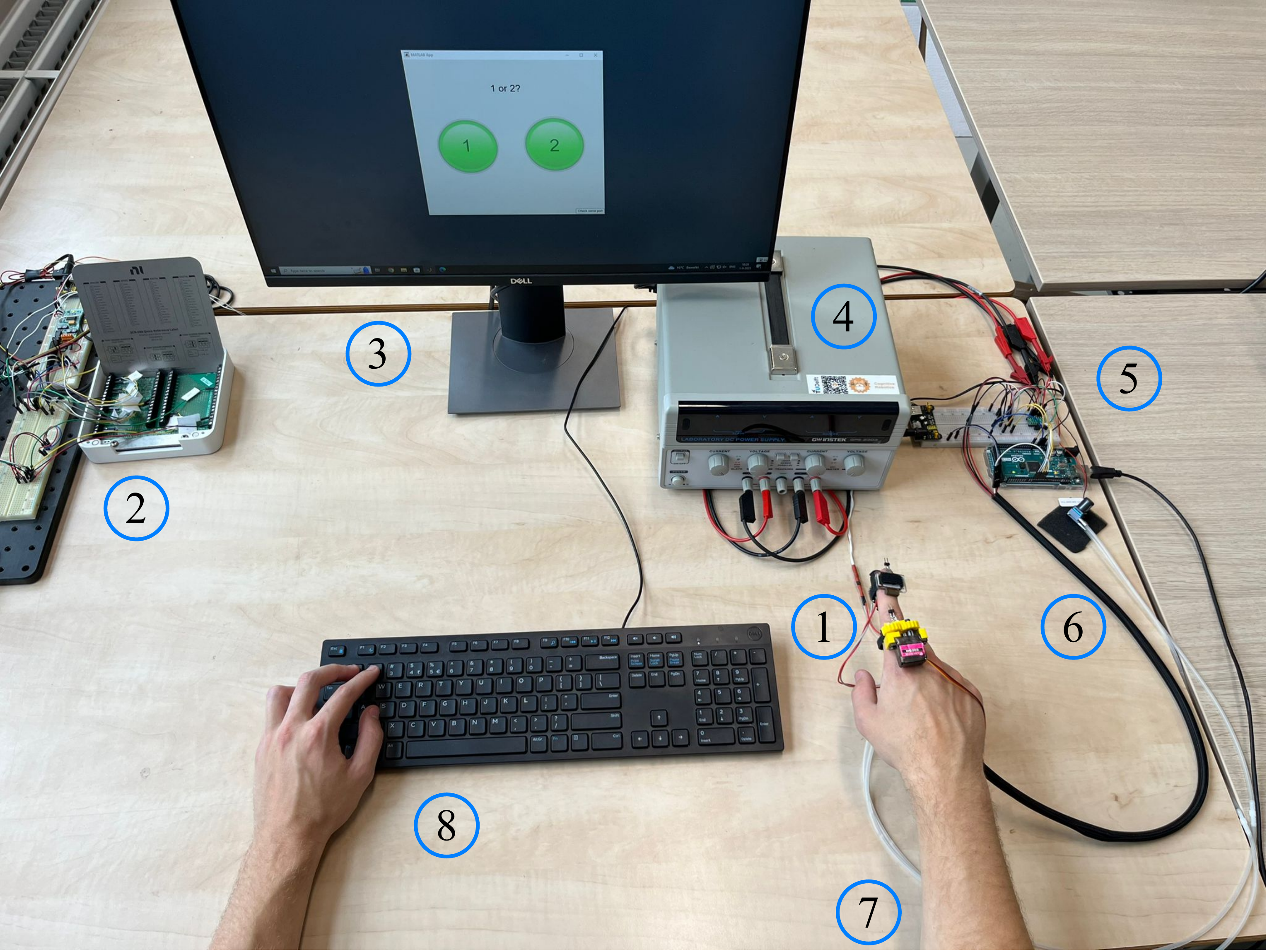}}\\
      \subfloat[ ]{%
       \includegraphics[width=\linewidth]{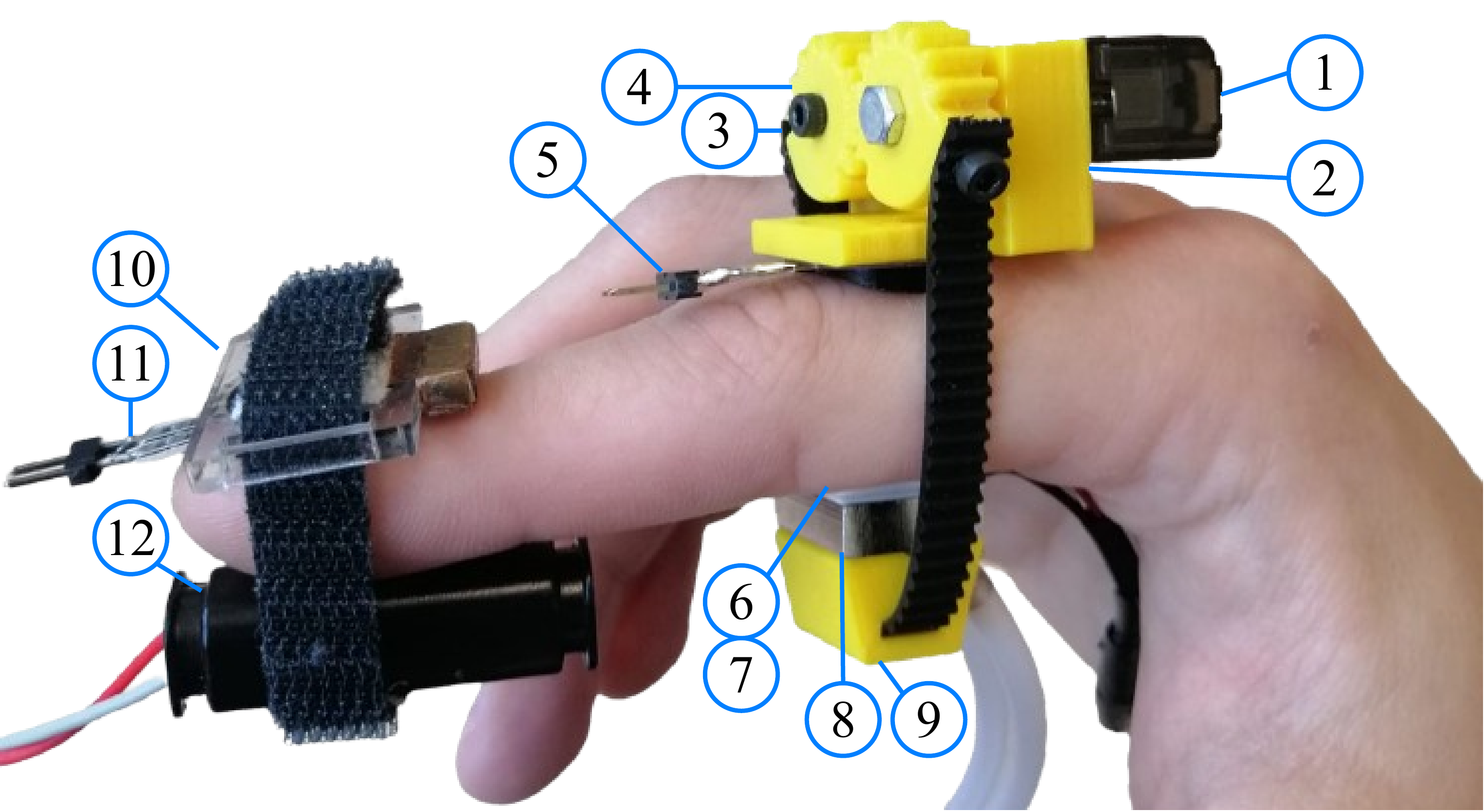}}
  \caption{(a) Experimental setup: 1. custom-designed multi-modal haptic device, 2. data acquisition board, 3. monitor displaying the GUI, 4. variable power supply, 5. Arduino, 6. water pump, 7. armrest, and 8. keyboard. (b) A closer look at the custom-designed haptic device: 1. servo motor, 2. servo motor housing, 3. servo belt, 4. spur gears, 5. force-resisting sensor, 6. NTC thermistor, 7. Peltier element, 8. water-cooling heat sink, 9. moving platform, 10. acrylic mounting plate, 11. force-sensing resistor, and 12. voice-coil actuator}
  \label{fig_components} 
\end{figure}

During the experiments, a participant sat in front of a monitor displaying a graphical user interface (GUI) and a keyboard (Fig.~\ref{fig_components}a). They wore a custom-designed multi-modal haptic device (Fig.~\ref{fig_components}b) displaying thermal and pressure stimuli to the proximal phalanx and vibrotactile stimuli on the fingerpad of the index finger of their right hand. 

Our device delivers thermal stimuli via a 15$\times
$15$\times$3.6~mm Peltier module (QC-31-1.0-3.9AS, QuickCool) mounted on the ventral side of the proximal phalanx. The Peltier temperature is regulated via a motor driver (DRV8833, Pololu) through a closed-loop PID controller by measuring temperature from a thermistor (GA10K3MCD1 NTC, TE Connectivity) placed between the skin and the Peltier module. The thermistor is connected to a microcontroller (Mega 2560
Rev 3, Arduino), which receives temperature data and handles the PID control of the Peltier module. This module sits on a water-cooled heat sink (MCX RAM, Alphacool), which is an effective method for regulating heat dissipation~\cite{gabardi2018development}~\cite{kodak2023feelpen}. A water pump (480-122, RS Pro) circulates the water through the heat sink with silicon hoses. The pump's input voltage was kept at 1.1~V throughout the entire experiment via a power supply (GPS-4303 DC Bench Power Supply, Gw Instek). Although fan cooling is also an option for regulating heat dissipation~\cite{gioioso2018using}~\cite{peters2023thermosurf}, fans are substantially larger compared to water cooling elements, which could hinder the freedom of movement, and water cooling has no moving mechanical parts. This assembly is placed on the skin via a 3D-printed mount connected to a motor belt. The two ends of the belt are connected to 3D-printed spur gears for adjusting the strapping pressure. The spur gears move in opposite directions, moving the platform up or down, driven by an MG90S servo motor controlled by the microcontroller. The applied pressure is measured via a force sensor (FSR06B, Ohmite) mounted under the gears. The vibrotactile cues were delivered via a voice coil actuator (Haptuator Mark II, Tactile Labs) mounted on the fingertip via a velcro strap~\cite{friesen2023perceived}. The signals for the actuator are generated via the data acquisition board and amplified by an audio amplifier (MIKROE-3077, Mikroe) with a 20 dB gain. The strapping force of the actuator is measured via a force sensor (FSR06B, Ohmite) mounted on the fingernail of the participant via the velcro strap, which sends force data to the data acquisition board. 

The participant rested their forearm on an armrest (Model 332020, Ergorest) to prevent fatigue. They also wore noise-canceling headphones (WH-1000XM3, Sony) playing pink noise to prevent audio bias and for playing sound cues. They gave their answers with their left hand through the keyboard.  

\subsection{Stimuli}
In our experiments, the vibrotactile stimulus was a sinusoidal signal with a frequency of 250~Hz applied on the fingertip via the voice-coil actuator. We chose this frequency because it lies within the sensitive frequency region (between 150 and 300~Hz) for human perception of vibrotactile stimuli~\cite{jones2008tactile}. Also, it allowed us to compare our findings with previous literature that used the 250~Hz vibration frequency~\cite{green1977effect}~\cite{oh2019effects}. The amplitude of the voice-coil actuator started at a high value (20~mV) and was adjusted during the experiments. The strapping force of the voice coil actuator was 0.5~N for all conditions to prevent setup slip from the hand during hand movements. 

We tested the effect of three thermal stimuli applied at two contact pressure values on the vibrotactile sensitivity, resulting in six experimental conditions (see Table~\ref{tab_conditions}). These were thermal stimuli of 40~\degree C (\emph{warm}), 18~\degree C (\emph{cold}), and 32~\degree C (\emph{neutral}), applied at contact pressures of either 0.5~N (\emph{low pressure}) or 2~N (\emph{high pressure}). We selected these values to avoid the noxious response reported for temperatures below 15~\degree C and above 45~\degree C ~\cite{ho2018material}. During the preliminary experiments, thermal stimuli below 18~\degree C and above 40~\degree C were perceived as painful by some participants; therefore, these values were selected as temperature limits. The pressure values were selected to keep functionality in mind, where the lowest pressure (0.5~N) corresponded to the strapping force, and 2~N was the highest pressure limit for comfort throughout one trial. 

\begin{threeparttable}[t]
\caption{Experimental conditions}
\label{tab_conditions}
    \setlength\tabcolsep{0pt}
\begin{tabular*}{\linewidth}{@{\extracolsep{\fill}} l cc cc @{}}
\toprule
Condition & Thermal & Pressure  \\ 
\midrule
Neutral - low pressure & 32~\degree C & 0.5 N \\
Neutral -  high pressure & 32~\degree C & 2 N \\
Cold - low pressure & 18~\degree C & 0.5 N \\ 
Cold - high pressure & 18~\degree C & 2 N \\ 
Warm - low pressure & 40~\degree C & 0.5 N \\ 
Warm - high pressure & 40~\degree C & 2 N  \\
\bottomrule
\end{tabular*}
\end{threeparttable}

\subsection{Experimental procedure}
Before the experiments, the participant washed their hands and used hand sanitizer. Then, the experimenter instructed the participant on how the experiment works. Afterward, the experimenter mounted the thermal and pressure module of the device to the participant's index finger and ensured that the mounting location was correct (see Fig.~\ref{fig_components}b). Then, the strapping force was increased by automatically adjusting the motor angle based on closed-loop force control until it reached the required pressure level. Then, the experimenter set the thermal condition until it reached the required temperature. Afterward, the experimenter mounted the vibrotactile module on the fingerpad of the participant and manually adjusted the velcro strap until achieving a strapping force of 0.5~N. Then, the participant placed their forearm on the armrest and put on the noise-canceling headphones. These calibration procedures took approximately 2-5 minutes. Afterward, the participant started the experiment without a training session by initiating the experiment by pressing the '2' key on the keyboard. Each participant conducted experiments with all conditions listed in Table~\ref{tab_conditions}. They completed the experiments in different random order.  

\begin{figure}[t]
    \centering
    \includegraphics[width=0.5\textwidth]{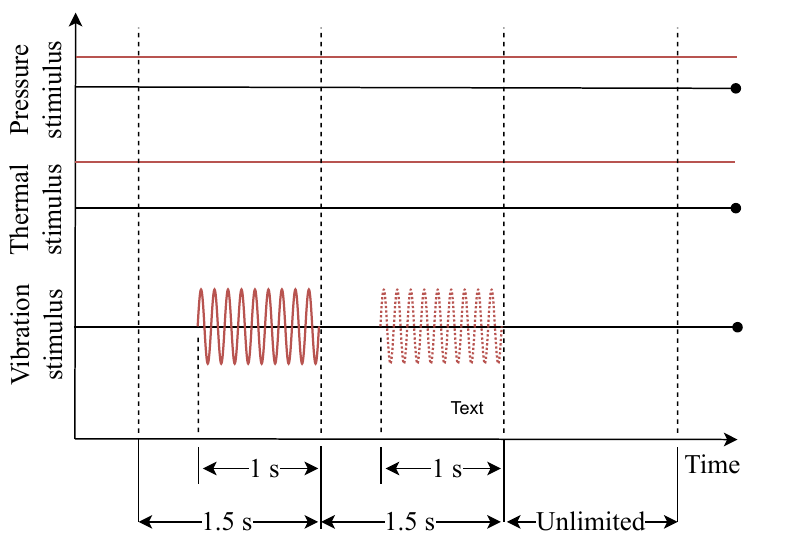}
    \caption{Stimuli timing diagram for the vibrotactile sensitivity experiments. A sinusoidal vibrotactile stimulus (250~Hz) with a duration of 1 second was either present in the first or the second interval randomly. Pressure and thermal stimuli were always present during the trial and kept constant throughout all repetitions. The participant had to choose in which of these two intervals the vibrotactile stimulus was present.}
    \label{fig_stimulitimingdiagram}
\end{figure}

We used a two-alternative-forced-choice method in our detection threshold experiments. The vibrotactile stimuli were presented with two temporal intervals signaled to the participants via the GUI as 1 and 2; only one of these intervals contained the test stimulus. The participant's task was to select the interval where they perceived a vibratory stimulus on their index fingertip. In each trial, the software randomized the interval during which the vibrotactile stimulus was present. A thermal stimulus was always present at the specified constant pressure during the entire session.

The participant was instructed to hold their index finger in the air and wait for a sound cue. Half a second later, the first interval played for 1 second. Then, the process was repeated for the second interval. After the second interval was completed, a different sound cue indicated it was time to select which of the two intervals contained the vibrotactile stimulus. The participant could answer by pressing the '1' or the '2' key on the keyboard, and they were indicated if their answer was correct or incorrect; check Fig.~\ref{fig_stimulitimingdiagram} for the stimuli timing diagram of the experiments.

\begin{figure}[t]
    \centering
    \includegraphics[width=0.49\textwidth]{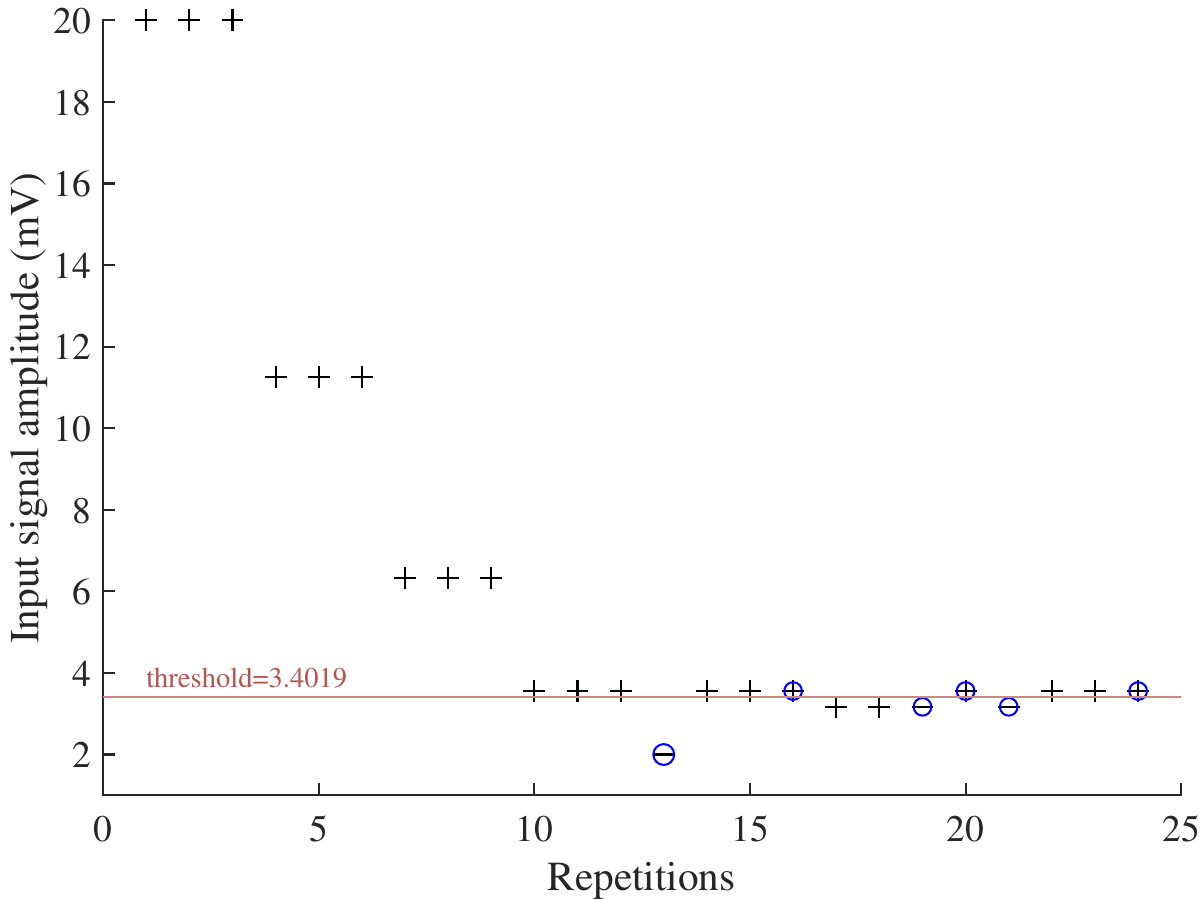}
    \caption{An example session of the adaptive three-up/one-down staircase method~\cite{zwislocki2001psychophysical}. The threshold value is calculated by averaging the last five reversals at the $\pm$1 dB range. Correct answers are represented by plus signs (+), and minus signs (-) represent incorrect answers. The circles (o) show reversals, and the threshold is indicated with a red line.}
    \label{fig_StaircaseExample}
\end{figure}

The amplitude of the input voltage signal to the voice-coil actuator was modulated using a three-up/one-down adaptive staircase method. This method obtains thresholds with a correct response probability of 75\%~\cite{zwislocki2001psychophysical}. The staircase started with an easily perceivable magnitude (20~mV). After three correct answers (not necessarily consecutive), the vibrotactile signal amplitude was decreased by 5~dB increments at the start, and after one wrong answer, this increment was reduced to 1~dB. When participants gave one wrong answer, the amplitude was increased by 1~dB. One complete trial was finished when the last five reversals remained in the $\pm1$~dB range, after which the mean of those five reversals revealed the detection threshold; see Fig.~\ref{fig_StaircaseExample} for an example session. Alternatively, to ensure participants' comfort, the trial stopped after 80 repetitions. When the session finished, the GUI indicated this, and a 10-minute resting period started. After six sessions (3 thermal $\times$ 2 pressure conditions), the experiment ended, resulting in a maximum duration of 2 hours. 

\section{Results}
\begin{figure}[t]
    \centering
    \includegraphics[width=0.49\textwidth]{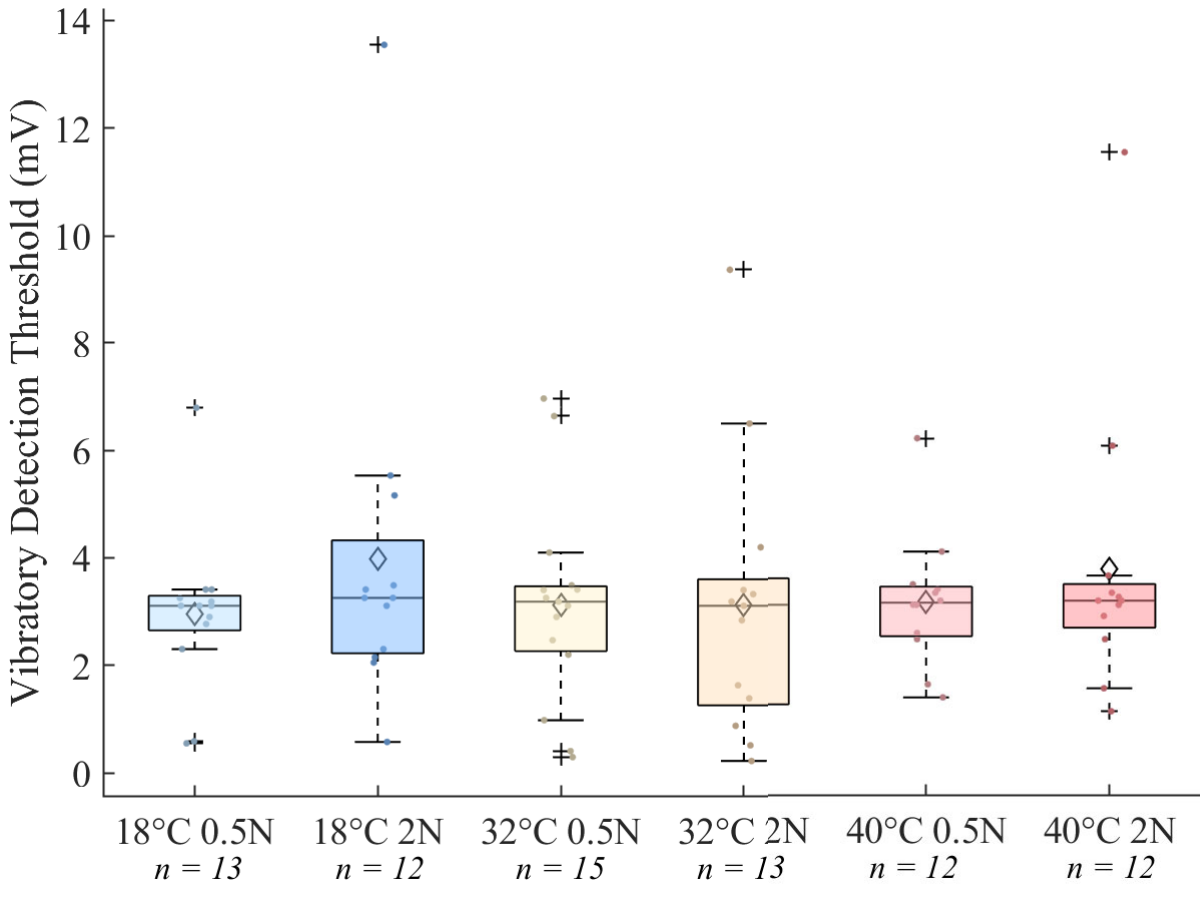}
    \caption{Boxplots of the vibrotactile detection thresholds. The results corresponding to each experimental condition are color-coded. The
center lines show the medians; box limits indicate the 25th and 75th percentiles. The whiskers extend to 1.5 times the interquartile range. Outliers are represented
by plus signs (+), and diamonds ($\diamond$) represent sample means. The points (.) show individual threshold values; the sample sizes (\emph{n}) are indicated under each boxplot.}
    \label{fig_VibratoryThreshold}
\end{figure}

The measured detection thresholds per condition are visualized in Fig.~\ref{fig_VibratoryThreshold}. The sessions in which the experiment was stopped due to reaching the maximum amount of repetitions (80) were excluded from the data analysis.

First, we applied Shapiro-Wilk tests~\cite{bensaida2023} to each distribution to check whether they distributed normally. For neutral-low pressure (Mean: 3.12, SD: 1.88), neutral-high pressure (Mean: 3.12, SD: 2.54), and warm-low pressure (Mean: 3.17, SD: 1.23) conditions, the Shapiro-Wilk test suggested that they follow a normal distribution (p~\textgreater{}~0.05). However, it indicated that cold-low pressure (Mean: 2.96, SD: 1.50), cold-high pressure (Mean: 3.99, SD: 3.29), and warm-high pressure (Mean: 3.78, SD: 2.73) conditions are not normally distributed (p~\textless{}~0.05). 

As not all collected data is normally distributed, we applied a two-way non-parametric Skillings-Mack test~\cite{skillings1981use}, suitable as an alternative to the Friedman test when data contains missing entries, to analyze the effects of relocated constant thermal and pressure stimuli on vibrotactile sensitivity on the fingertip. We evaluated the effect of relocated thermal stimuli by taking three thermal conditions as treatments and two pressure conditions as blocks. Our results showed that relocated thermal stimuli did not significantly affect the vibratory detection thresholds (T(2) = 0.23, p = 0.893). % T = 0.22702, df = 2
Similarly, we evaluated the effect of relocated pressure by taking two pressure conditions as treatments and three thermal conditions as blocks. Our results showed that relocated pressure stimuli also did not significantly affect the vibratory detection thresholds (T(1) = 0.14, p = 0.7106). % T = 0.13768, df = 1

\section{Discussion}
In this study, we sought to understand the impact of thermal stimuli applied to the proximal phalanx of the finger on the vibrotactile sensitivity of the fingertip. For this aim, we designed a novel haptic device capable of delivering controlled thermal stimuli to the proximal phalanx at varying pressure levels while administering vibrotactile stimuli to the index fingertip. We then measured the detection thresholds of 15 participants for 250~Hz vibration stimuli for warm and cold stimuli at high and low-pressure conditions.   

Our results showed no significant effect of relocated thermal stimuli---regardless of the applied pressure level---on vibratory detection thresholds measured on the fingertip  (see Fig.~\ref{fig_VibratoryThreshold}). This finding may be due to thermal adaptation, a phenomenon in which participants become less responsive to \emph{constant} thermal stimuli due to continuous exposure over time~\cite{ho2018material}. Interestingly, participants in our study reported that thermal stimuli felt most intense during initial calibration. Thus, focusing on vibrotactile stimuli might have become less demanding during thermal conditions when the repetition number increased. It is also possible that thermal and vibrotactile stimuli may be processed independently~\cite{park2022preliminary} and do not affect each other when applied at different skin locations. To test this hypothesis, we plan to repeat the same experiments in future work while initializing the thermal stimulus \emph{dynamically} in each trial.

We observed large variabilities between participant responses. For example, participant 4, who had the lowest vibration threshold at neutral condition, could not reach the threshold values when cold and warm conditions. For this participant, thermal conditions caused too much distraction, resulting in the inability to finish the experiment within 80 repetitions. In contrast, participant 10 had the highest overall thresholds for all conditions. However, they were affected more by pressure than other participants, showing a substantial difference between low and high-pressure conditions. Participant 3 had issues with converging for the warm conditions, while they found cold ones refreshing. Previous work also mentions this high degree of variation in perception of vibrotactile, thermal, and pressure stimuli~\cite{fagius1981variability}. They state that not only personal qualities cause variability but also distraction, motivation, fatigue, and anxiety due to testing participants' abilities. However, this variability is most influential when studying individual participants and has less effect on the overall comparison between condition groups. 

Despite carefully designed experimental setup and procedures, our work had several limitations that could affected our results. For example, the selected temperature and pressure levels were conservative and within comfortable ranges. More extreme values just above the pain threshold could alter the vibrotactile sensitivity~\cite{apkarian1994gate}. Moreover, device ergonomics limited applied pressure ranges. Attachment forces lower than 0.5~N caused the haptic device to slip when moving around. Adding an external mount could prevent this slippage and open up the possibility of reducing the attachment force even further. However, this addition introduces a second pressure point to the assembly, which is why it was excluded from the current design.

As far as we know, this study is the first attempt to investigate the interplay between cross-modal tactile stimuli when applied to distinct skin locations. Our preliminary findings suggest that thermal and vibrotactile stimuli do not impede each other's perception when administrated at different sites. Delivering thermal stimuli through a wearable ring that refrains from obstructing fingertips (Fig.~\ref{fig_relocation}) without impeding the sensation at the fingertips can unlock novel avenues for haptic feedback in various applications, particularly in telerobotics, where finger sensitivity is paramount, such as surgical procedures. Furthermore, we hope our wearable device design is valuable to designers and engineers.

\section*{Acknowledgments}
The authors thank Bence Kodak and 
Jagan K. Balasubramanian for their insights and suggestions, Jacques Brenkman and Maurits Pfaff for their advice on the experimental setup. 
\newpage
\bibliographystyle{IEEEtran}
\bibliography{bibliography}
\end{document}